\documentclass[a4paper]{article}
\usepackage[english]{babel}
\usepackage[utf8]{inputenc}
\usepackage[T1]{fontenc}
\usepackage{authblk}
\usepackage{graphicx}
\usepackage{amsmath,mathrsfs,dsfont}
\usepackage{amsthm}
\usepackage[all]{xy}
\usepackage{graphics}
\usepackage{tensor}
\usepackage{verbatim}
\usepackage{nicefrac}
\usepackage{amsfonts}
\usepackage{amssymb,latexsym}
\usepackage{color}
\usepackage{mathrsfs,dsfont}
\usepackage{appendix}
\usepackage{slashed}
\usepackage{cite}
\usepackage{url}
\usepackage{hyperref}
\usepackage{comment}
\newcommand{\Tr}{\mathrm{Tr}}
\newcommand{\ba}{\begin{eqnarray}}
\newcommand{\ea}{\end{eqnarray}}

\begin{document}

\title{On a non-geometric approach to noncommutative gauge theories}
\author[1]{Guilherme R. G. Barrocas\thanks{guibarrocas@gmail.com}}
\author[1,2]{Aleksandr Pinzul\thanks{apinzul@unb.br}}
\affil[1]{Universidade de Bras\'{\i}lia, Instituto de F\'{\i}sica 70910-900, Bras\'{\i}lia, DF, Brasil}
\affil[2]{International Center of Physics C.P. 04667, Brasilia, DF, Brazil}
\date{}
\maketitle

\begin{abstract}
In this work, we generalize the non-geometrical construction of gauge theories, due to S.~Deser, to a noncommutative setting. We show that in a free theory, along with the usual local N\"{o}ther current, there is another conserved current, which is non-local. Using the latter as a source for self-interaction, after a well-defined consistency procedure, we arrive at noncommutative gauge theories. In the non-abelian case, the standard restriction, namely that the theory should be $U(N)$ in the fundamental representation, emerges as a consequence of the requirement that the non-local current be Lie algebra valued.
\end{abstract}

\section{Introduction}

Field theories on noncommutative space-time were introduced as earlier as, at least, 1947 \cite{Snyder:1946qz}, as an attempt to regularize the UV behaviour. However, the real surge of interest in the topic came after it was shown in \cite{Seiberg:1999vs} that noncommutative gauge theories naturally appear in a specific limit (called the decoupling limit) of superstring theory.

The standard approach to noncommutative gauge theories is based on deforming the geometry of a commutative gauge theory into a consistent noncommutative counterpart. In this sense, the underlying commutative geometry of a gauge theory is already assumed and this is what we call the geometric approach. The consistency of the deformation imposes some severe restrictions on the resulting noncommutative gauge theories, see \cite{Douglas:2001ba,Szabo:2001kg} for reviews of the standard approach. To avoid these restrictions, one has to resort to a construction based on enveloping algebras, which substantially alters the algebraic structure of the theory, see \cite{Jurco:2000ja} for a review.

In \cite{Deser:1969wk}, a purely field-theoretic (meaning: not relying on the geometry of a gauge field) approach to constructing non-abelian gauge fields was suggested. This approach was motivated by a similar approach to gravity \cite{Feynman:1996kb}, which in \cite{Deser:1969wk} was also dramatically simplified. The main idea is to construct a consistent self-interacting theory starting with a free one. This is a kind of ``bootstrap'' construction, where gauge symmetry emerges as the result of some well-defined consistency procedure rather then through gauging a global symmetry. The outcome of such a construction is the usual Yang-Mills theory, but the local (gauge) symmetry appears to be accidental rather than being built-in from the outset.

In this paper, we adopt this approach to study the possibility of such a non-geometric bootstrap for noncommutative gauge theory. This is intrinsically non-trivial, because free theories (which are the staring point of the construction) on noncommutative Moyal space-time (i.e., with noncommutativity introduced via a star-product, see section \ref{NC} for details) are exactly the same as in the commutative case. We show that there is a possibility of having two different consistent self-interacting theories, starting from the same free theory, and this is due to the fact that there are two conserved currents, one local and one non-local. Using the former as a source of self-interaction leads to the usual commutative Yang-Mills theory, while the theory based on the latter, non-local, current results in a noncommutative gauge theory. In the $U(1)$ case, the construction is particularly non-trivial: while there is no conserved local current, there is a non-zero non-local one. Thus, while the commutative theory is forced to remain free, its noncommutative counterpart becomes an interacting noncommutative $U(1)$ gauge theory. The $U(1)$ case illustrates all the main steps/details of our approach and its generalization to a non-abelian case results in the same restrictions on the gauge group and the allowed representations that are know from the standard approach.

The plan of the paper is as follows. In section \ref{Deser}, we review the Deser's approach \cite{Deser:1969wk} in the form that will be suitable for noncommutative generalization. Section \ref{NC} gives such a generalization. In particular, in subsection \ref{U1}, we provide the detailed construction of the noncommutative $U(1)$ theory, which contains all the main steps of our approach. Subsection \ref{Non-abelian} discusses the non-abelian generalization. We conclude with discussion of the obtained results and of some remaining questions.

\section{Deser's approach}\label{Deser}

Here we describe the non-geometric construction of non-abelian gauge fields, first done in \cite{Deser:1969wk}.

Let $G$ be a Lie group, whose Lie algebra $\mathfrak{g}$ is reductive, compact and not necessarily semi-simple, that is, the center $\mathcal{Z}\subset \mathcal{G}$ may be non-empty, for example, $G=U(N)$. These conditions guarantee that the generators of $\mathfrak{g}$, $T^a,\ a=\overline{1,dim\,\mathfrak{g}}$, are Hermitian, i.e., the representation of $G$ is unitary\footnote{Even if a finite dimensional representation is not unitary, one can always ``unitarize'' it by the Weyl’s unitarian trick. This is guaranteed by the existence of the Haar measure and compactness, see, e.g., \cite{Hall:371445}.}, and can be chosen such that
\ba\label{generators}
[T^a, T^b]=ic^{ab}_{\ \ c} T^c ,\ \Tr\,( T^a T^b ) = \delta^{ab},
\ea
where the structure constants are completely antisymmetric and there is no distinction between lower and upper indices (for example, $c^{ab}_{\ \ c} = c^{abc} = c_{abc}$).\footnote{We use the ``physicist'' choice when the generators are Hermitian, hence $i$ in front of the structure constants in Eq.(\ref{generators}).}

Take $dim\,\mathfrak{g}$ copies of fields $F_a^{\mu\nu}$ and $A_{a\mu}$, $a=\overline{1,dim\,\mathfrak{g}}$ and define $\mathbf{F}^{\mu\nu}:=F_a^{\mu\nu} T^a$, $\mathbf{A}^{\mu}:=A_a^{\mu} T^a$. Consider the following quadratic action (free dynamics) for these fields (we use the signature $(-,+,+,+)$)
\ba\label{action_free}
S_0 = -\frac{1}{2}\int d^4 x\, \Tr\left[\mathbf{F}^{\mu\nu}(\partial_\mu \mathbf{A}_\nu - \partial_\nu \mathbf{A}_\mu) - \frac{1}{2} \mathbf{F}^{\mu\nu}\mathbf{F}_{\mu\nu}\right].
\ea

The equations of motion that follow from (\ref{action_free}) are easily found
\ba\label{eom_comm}
\mathbf{F}^{\mu\nu}&:& \mathbf{F}_{\mu\nu} = \partial_\mu \mathbf{A}_\nu - \partial_\nu \mathbf{A}_\mu , \label{eomA_comm}\\
\mathbf{A}_{\nu}&:& \partial_\mu \mathbf{F}^{\mu\nu} = 0 \label{eomB_comm}.
\ea
Due to (\ref{generators}), in the commutative pure gauge theory (i.e., without matter) case, nothing really depends on the choice of the representation of the $T^a$'s. We will see how drastically the situation changes in the noncommutative case.

By using (\ref{eomA_comm}), which are just constraints, and plugging them back into (\ref{action_free}), we obtain
\ba\label{action_free_F}
S_0 = -\frac{1}{4}\int d^4 x\, \Tr\left[\mathbf{F}^{\mu\nu}\mathbf{F}_{\mu\nu}\right] \equiv -\frac{1}{4}\int d^4 x\, {F}_a^{\mu\nu}{F}_{a\mu\nu},
\ea
i.e., $dim\,\mathfrak{g}$ copies of non-interacting $U(1)$ gauge fields $A_{a\mu}$. Then (\ref{action_free}) is just a Palatini-like first order formulation of this free theory. For us, (\ref{action_free}) will be more convenient.

The action (\ref{action_free}) is invariant under \textbf{global} adjoint $G$-transformations
\ba\label{global}
\mathbf{F}^{\mu\nu} \mapsto g\mathbf{F}^{\mu\nu}g^{-1},\  \mathbf{A}_{\mu} \mapsto g\mathbf{A}_{\mu}g^{-1},\ g\in G.
\ea
We now construct the N\"{o}ther current associated with the symmetry (\ref{global}) by using the standard trick. Though it is standard textbook material, because it will be very important in the noncommutative case, we will briefly summarize the procedure: By \textbf{formally} promoting $g\in G$ to be a group-valued function, the action is no longer invariant under (\ref{global}), but rather its infinitesimal variation will take the form
\ba\label{variation}
\delta S_0 = S_0^g - S_0 = \int d^4 x\, \Tr (\mathbf{j}_0^{\mu}\partial_\mu {\omega}) + \mathcal{O}(\omega^2)\ ,
\ea
where $g = e^{i\omega} \equiv e^{i\omega_a T^a}$, $\omega_a$ being a set of arbitrary functions. As usual, the action is stationary on-shell, which means that $\delta S_0$ must vanish under the (infinitesimal form of the) variations (\ref{global}) even when $g$ is non-constant, provided the equations of motion hold. Since in (\ref{variation}) only $\omega$ is arbitrary and not $\partial_\mu \omega$, we conclude that the current $\mathbf{j}_0^{\mu}$ must be conserved on the equations of motion (\ref{eomA_comm},\ref{eomB_comm}), $\partial_\mu\mathbf{j}_0^{\mu} = 0$.

Comment (important): Promoting $g$ to be a function rather than a constant element of $G$ has nothing to do with gauging! This is merely a trick to get a N\"{o}ther current. The action still remains invariant only under {global} $G$-transformations.

Carrying out the procedure described above, the variation of (\ref{action_free}) is
\ba\label{variation1}
\delta S_0 = \int d^4 x\, \Tr \bigl([\mathbf{A}_\mu , \mathbf{F}^{\mu\nu}]g^{-1}\partial_\nu g\bigr)=i\int d^4 x\, \Tr \bigl([\mathbf{A}_\mu , \mathbf{F}^{\mu\nu}]\partial_\nu {\omega}\bigr) + \mathcal{O}(\omega^2).
\ea
Thus the conserved N\"{o}ther current is
\ba\label{current}
\mathbf{j}_0^\mu = iq[\mathbf{A}_\nu , \mathbf{F}^{\nu\mu}],\ \mathrm{or}\ j_{0a}^\mu = iq\Tr ([\mathbf{A}_\nu , \mathbf{F}^{\nu\mu}] T^a)\equiv - q c^{abc} A_{b\mu}F_c^{\mu\nu},
\ea
where $q$ is a ``coupling constant''. By using (\ref{eomA_comm},\ref{eomB_comm}), it is an easy exercise to verify the conservation of this current (we will do this explicitly in the less trivial noncommutative case, see the next section).

From (\ref{current}), it is clear that all the components of the current, corresponding to the generators of the center, $\mathcal{Z} \subset\mathcal{G}$, vanish, as expected. This will be crucial below, when we will be constructing a self-interacting theory: the $U(1)$ fields corresponding to these generators will remain free in agreement with the general result \cite{Deser:1963zzc}.

Now, we construct a consistent self-interacting theory by using the current (\ref{current}) as the source for the free equation of motion (\ref{eomB_comm}). Although each component of the current (\ref{current}) is separately conserved, the argument from \cite{Deser:1963zzc} can be used to argue that one has to use the full current to get a consistent interaction.\footnote{Actually, one can use a ``sub-current'' corresponding to a non-abelian sub-group, $G_1 \subset G$. Then only the fields $F_a^{\mu\nu}$, $A_{a\mu}$, where $a$ is such that $T^a \in \mathcal{G}_1$ will become interacting leaving the rest to remain free.} Then the equation (\ref{eomB_comm}) is modified to
\ba\label{eom_source}
\partial_\mu \mathbf{F}^{\mu\nu} = -\mathbf{j}_0^\nu ,
\ea
while the constraint (\ref{eomA_comm}) remains without a change.

The issue with (10) is that while the divergence of the left-hand side vanishes identically, i.e., not only dynamically ($\partial_\nu\partial_\mu \mathbf{F}^{\mu\nu} \equiv 0$ due to antisymmetry of $\mathbf{F}^{\mu\nu}$), the right-hand side is conserved only after the old dynamics (\ref{eomA_comm},\ref{eomB_comm}) is imposed. Of course, this is due to the fact that (\ref{eomA_comm}) is the constraint that followed from the action (\ref{action_free}) and the new dynamics given by (\ref{eom_source}) does not follow from that action. So, we must modify the action in such a way as to get (\ref{eom_source}) as the new equations of motion. The modification is rather straightforward:
\ba\label{action_inter}
S = S_0 +\frac{1}{2}\int d^4 x\, \Tr \bigl( \mathbf{j}_0^\mu \mathbf{A}_\mu \bigr) = -\frac{1}{2}\int d^4 x\, \Tr\left[\mathbf{F}^{\mu\nu}(\partial_\mu \mathbf{A}_\nu - \partial_\nu \mathbf{A}_\mu + iq [\mathbf{A}_\mu , \mathbf{A}_\nu] ) - \frac{1}{2} \mathbf{F}^{\mu\nu}\mathbf{F}_{\mu\nu}\right] .
\ea
The slightly less intuitive coefficient of $\frac{1}{2}$ is due to the fact that the current depends on $\mathbf{A}$, but the main argument is that the resulting equations of motion are what we need:
\ba\label{eom_comm1}
\mathbf{F}^{\mu\nu}&:& \mathbf{F}_{\mu\nu} = \partial_\mu \mathbf{A}_\nu - \partial_\nu \mathbf{A}_\mu + iq [\mathbf{A}_\mu , \mathbf{A}_\nu] , \label{eomA_comm1}\\
\mathbf{A}_{\nu}&:& \partial_\mu \mathbf{F}^{\mu\nu} = -i q [\mathbf{A}_\mu , \mathbf{F}^{\mu\nu}]\equiv -\mathbf{j}^\nu \label{eomB_comm1}.
\ea
Thus we see that while we get (\ref{eom_source}) as the new equation of motion, the old constraint (\ref{eomA_comm}) gets modified.

Of course, because (\ref{eomA_comm1},\ref{eomB_comm1}) are the standard Yang-Mills equations, it is known that they are already consistent, and the procedure terminates after the first step\footnote{In \cite{Deser:1969wk}, it is shown that should we have started with the second order formulation (\ref{action_free_F}), the ``consistency'' procedure would be one step longer with the same final result. The situation is rather different in the gravity case: while the first order formulation terminates after two steps, producing the usual GR action, the second order formulation of the free theory, i.e., the Fierz-Pauli theory, will lead to a consistent self-interacting theory only after infinite number of steps, see, e.g., \cite{Feynman:1996kb}.}, but it is instructive to verify it explicitly, demonstrating that the process of constructing a consistent self-interacting theory really terminates. Towards this end, notice that because the current $\mathbf{j}_0^\mu$ also transforms in the adjoint representation under global transformations and the interaction term in (\ref{action_inter}) does not contain derivatives, it is invariant under local transformations, too. Then we immediately conclude that the new N\"{o}ther current will be exactly the same as the old one:
\ba\label{current_full}
\mathbf{j}^\mu = \mathbf{j}_0^\mu = iq [\mathbf{A}_\mu ,\mathbf{F}^{\mu\nu}] .
\ea
This means that it is conserved on the new dynamics, too (again, in the noncommutative case, we will check this explicitly)
\ba\label{conserve}
\partial_\mu \mathbf{j}^\mu = 0\ \mathrm{on}\ (\ref{eomA_comm1},\ref{eomB_comm1}).
\ea
This observations proves the consistency of (\ref{eomB_comm1}).

Because of its importance and in view of the future generalization to the noncommutative case, we briefly recall how one sees the gauge invariance of the self-interacting theory (\ref{action_inter},\ref{eomA_comm1},\ref{eomB_comm1}). As usual, by re-writing (\ref{eomB_comm1}) in the form
\ba\label{covariant}
\partial_\mu \mathbf{F}^{\mu\nu} +i q [\mathbf{A}_\mu , \mathbf{F}^{\mu\nu}] =: \mathbf{D}_\mu^A \mathbf{F}^{\mu\nu} = 0,
\ea
we introduce the covariant derivative (in the adjoint representation), $\mathbf{D}_\mu^A$. It is a routine check that under the gauged version of (\ref{global}):
\ba\label{local}
\mathbf{A}_\mu \mapsto g \mathbf{A}_\mu g^{-1} + \frac{i}{q}\partial_\mu g g^{-1},
\ea
$\mathbf{F}^{\mu\nu}$ and $\mathbf{D}_\mu^A$ transform covariantly, i.e.,
\ba\label{covariant1}
\mathbf{F}^{\mu\nu} \mapsto g \mathbf{F}^{\mu\nu} g^{-1},\ \ g \mathbf{D}_\mu^A \mathbf{F}^{\mu\nu} g^{-1}
\ea
even if $g\in G$ is not constant. This demonstrates the \textbf{emerged} gauge invariance of the consistent self-interacting theory.

We want to stress one more time that by the construction, the action (\ref{action_inter}) and the equations of motion that follow from it, (\ref{eomA_comm1},\ref{eomB_comm1}), are guaranteed to be invariant only with respect to the global transformations (\ref{global}). The fact that they are also invariant with respect to the local $G$-transformations, (\ref{local}), in this approach, is \textbf{accidental}, i.e., the gauge invariance is emergent instead of being built-in from the very beginning. This is the most important point of this approach (and the main difference from the geometric approach with the built-in gauge symmetry), which we want to adopt for the non-geometric construction of noncommutative gauge theories, where the geometry is not always under an easy control.

\section{Noncommutative gauge theory}\label{NC}
Now we would like to adopt the approach from the previous section to the noncommutative setup. We begin with the $U(1)$ case, which is somewhat simpler technically but particularly striking: it shows the effectiveness of the approach, because it demonstrates how noncommutativity allows for consistent self-interaction even in the abelian case. Also, it already contains all the non-trivial steps of a non-commutative generalization. Later, we extend the construction to the non-abelian case, highlighting only the new features due to non-abelianity.

\subsection{Noncommutative $U(1)$}\label{U1}
Consider a noncommutative algebra $(\mathcal{A}, *)$, where $\mathcal{A}$ is a linear space of some ``nice'' functions (for us, it will be enough to think that the elements of $\mathcal{A}$ vanish at infinity sufficiently fast, e.g., that they are Schwartz functions) and $*$ is the Moyal star-product \cite{Groenewold:1946kp,Moyal:1949sk}, a noncommutative product defined for any two elements $a,b\in\mathcal{A}$ as
\ba\label{Moyal}
(a*b)(x)&:=& \sum\limits_{n=0}^{\infty} \frac{i^n}{n!}\theta^{\mu_1 \nu_1}\cdots \theta^{\mu_n \nu_n}\bigl( \partial_{\mu_1} \cdots \partial_{\mu_n} a(x) \bigr) \bigl( \partial_{\nu_1} \cdots \partial_{\nu_n} b(x) \bigr)\nonumber \\
&=& \Bigl( e^{i\theta^{\mu\nu}\partial_{\mu}^x \partial_{\nu}^y} a(x) b(y) \Bigr)_{x=y} \equiv a(x)e^{i\overleftarrow{\partial}_{\mu} \theta^{\mu\nu}\overrightarrow{\partial}_{\nu}} b(x),
\ea
where $\theta^{\mu\nu}$ is an antisymmetric constant matrix, with $||\theta||$ being the scale of noncommutativity.

The star-product is associative, $(a*b)*c=a*(b*c)=a*b*c$, and the most important, for us, property of the Moyal star is that the deformation of the commutative product is always a total derivative
\ba\label{deformation}
(a*b)(x) = a(x)b(x) + \partial_{\mu_1} \Bigl( \sum\limits_{n=1}^{\infty} \frac{i^n}{n!}\theta^{\mu_1 \nu_1}\cdots \theta^{\mu_n \nu_n}\bigl( \partial_{\mu_2} \cdots \partial_{\mu_n} a(x) \bigr) \bigl( \partial_{\nu_1} \cdots \partial_{\nu_n} b(x) \bigr)\Bigr).
\ea
This property allows us to always ``integrate out'' one star (under the suitable boundary conditions for the elements of $\mathcal{A}$, which we always assume)
\ba\label{cyclic}
\int (a*b)\,d^4 x = \int (ab)\,d^4 x ,
\ea
which, in turn, implies cyclicity of the $*$-product under the integral (which then plays the role of the trace on $\mathcal{A}$).

Now consider unitary elements in $\mathcal{A}$, $u\in U(\mathcal{A})\subset\mathcal{A}$ and $u*u^\dagger = u^\dagger * u = \mathds{1}$, where $\dagger$ is essentially the complex conjugation and $(a*b)^\dagger = b^\dagger * a^\dagger$ for all $a,b \in \mathcal{A}$, which immediately follows from the definition (\ref{Moyal}). It is easy to see that any such unitary element can be written as a star-exponential:
\ba\label{starEXP}
u= e_*^{i\omega}:= \sum\limits_{n=0}^{\infty} \frac{(i)^n}{n!} \underset{n\ times}{\underbrace{\omega * \cdots * \omega}},
\ea
for some real, $\omega = \omega^\dagger$, element of $\mathcal{A}$.

As in the previous section, we start with the first-order formulation of just one copy of a free theory
\ba\label{commU1}
S_0 = -\frac{1}{2}\int d^4 x \left[F^{\mu\nu}(\partial_\mu A_\nu - \partial_\nu A_\mu) - \frac{1}{2} F^{\mu\nu}F_{\mu\nu}\right].
\ea
The usual equations of motion are
\ba
F^{\mu\nu}&:& F_{\mu\nu} = \partial_\mu A_\nu - \partial_\nu A_\mu , \label{eomA}\\
A_{\nu}&:& \partial_\mu F^{\mu\nu} = 0 \label{eomB}.
\ea

Because for a constant $u$, the transformations (\ref{global}) are trivial in the $U(1)$ case, there is no non-trivial global symmetry, hence there is no conserved (local) N\"{o}ther current.\footnote{In the presence of matter, there is global $U(1)$ symmetry, which leads to the conserved ``electric'' charge, but here we consider a pure gauge theory.} (As we explicitly commented after (\ref{current}), of course, this is the case not only for $U(1)$ global symmetry but for any abelian group.) Treating $F^{\mu\nu}$ and $A_\mu$ as elements of the noncommutative algebra $\mathcal{A}$, consider the following transformations
\ba\label{ncU1}
\forall u\in U(\mathcal{A}),\ \left\{
\begin{array}{ccc}
  F^{\mu\nu} & \mapsto & u*F^{\mu\nu}*u^\dagger =: F^{\mu\nu}_u \\
  A^{\mu} & \mapsto & u*A^{\mu}*u^\dagger =: A^\mu_u
\end{array}
\right. .
\ea
Clearly, if $u$ is a constant unitary element of $\mathcal{A}$, (\ref{ncU1}) will become trivial and the action will be trivially invariant, as above. Now we follow the N\"{o}ther procedure outlined in the previous section. So, consider the case of non-constant $u$. Then, the action (\ref{commU1}) evaluated for $F_u$ and $A_u$, $S_0^u$, will be equal
\ba\label{commU1u}
S_0^u = S_0 - \int d^4 x \left[ A_\nu * F^{\mu\nu} - F^{\mu\nu}*A_\nu \right] *u^\dagger * \partial_\mu u .
\ea
In deriving (\ref{commU1u}), we used (multiple times) the cyclic property of the Moyal star under integral. So, when $u = const$, we trivially have $S_0^u = S_0$, as it should be. Taking $\omega$ in (\ref{starEXP}) to be a non-constant element of $\mathcal{A}$, we have at the leading order in $\omega$
\ba
%&& F^{\mu\nu}_u - F^{\mu\nu} =: \delta F^{\mu\nu} = i[\omega, F^{\mu\nu}]_* ,\label{epsilon1}\\
%&& A^{\mu}_u - A^{\mu} =: \delta A^{\mu} = i[\omega, A^{\mu}]_* ,\label{epsilon2}\\
&& u^\dagger * \partial_\mu u = i\partial_\mu \omega \label{epsilon3}.
\ea
Using this in (\ref{commU1u}), it takes the form of the equation (\ref{variation}). Then we conclude that the following \textbf{non-local} current is conserved
\ba\label{NLcurrent1}
j_0^\mu = i q [A_\nu , F^{\nu\mu}]_*,\ \ \partial_\mu j_0^\mu = 0\ \ \mathrm{on\ the\ e.o.m.},
\ea
where $[a,b]_* = a*b - b*a$ and $q$ is again a ``coupling constant''.\footnote{This is an analogue of the non-local energy momentum tensor in the presence of noncommutativity, studied in \cite{Pinzul:2011di}. This object should be relevant if one tries to adopt the method of the present work to construct noncommutative gravity.}
There is no contradiction with the stated above fact that there is no N\"{o}ther current for the global $U(1)$. As we already mentioned, it is true only for a local current, while (\ref{NLcurrent1}) is clearly non-local due to the non-locality of the star-product. When the noncommutative parameter, $||\theta ||$, is sent to zero (or, rather, if the characteristic scale, on which fields fluctuate, is much bigger than $||\theta ||^{1/2}$), $j_0^\mu \rightarrow 0$.

Though the derivation of the current (\ref{NLcurrent1}) formally followed the standard procedure of deriving a N\"{o}ther current outlined in the previous section, due to the non-locality of the obtained result, it is useful to explicitly check that $j_0^\mu$ is conserved on the e.o.m. (\ref{eomA},\ref{eomB}):
\ba\label{check1}
\frac{1}{iq}\partial_\mu j_0^\mu =  [\partial_\mu A_\nu , F^{\nu\mu}]_* +  [A_\nu , \partial_\mu F^{\nu\mu}]_* \overset{e.o.m.}{=} -\frac{1}{2}[F_{\mu\nu} , F^{\mu\nu}]_* \equiv 0.
\ea

Following Deser, we want to construct a consistent self-interacting theory now sourced by the non-local current (\ref{NLcurrent1}):
\ba\label{modification1}
\partial_\mu F^{\mu\nu}= - j_0^\nu \equiv -i q [A_\mu , F^{\mu\nu}]_* ,
\ea
where, again, the relation between $F$ and $A$ remains the same (\ref{eomA}). Exactly as in the commutative case, because $\partial_\nu \partial_\mu F^{\mu\nu}=0$ identically, i.e., without evoking the equations of motion and the divergence of the right hand side is not zero on the new equations of motion, the old current $j_0^\mu$ is not conserved on the new e.o.m.. Then, as in the commutative case, (\ref{modification1}) is not consistent and we have to run the ``consistency procedure''. We modify the original action (\ref{commU1}) as in (\ref{action_inter}) but now inserting a star-product (though, as we mentioned, it could be integrated out, we will need it to do cyclic permutations):
\ba\label{action1}
S = S_0 + \frac{1}{2}\int d^4 x\, (j_0^\mu * A_\mu ) = - \frac{1}{2}\int d^4 x  \left[ F^{\mu\nu}(\partial_\mu A_\nu - \partial_\nu A_\mu +i q [A_\mu , A_\nu]_*) - \frac{1}{2} F^{\mu\nu}F_{\mu\nu} \right],
\ea
where we again used cyclicity of the star-product and, at the end, integrated out one star. The new equations of motion that follow from (\ref{action1}) are easily found
\ba
F^{\mu\nu}&:& F_{\mu\nu} = \partial_\mu A_\nu - \partial_\nu A_\mu + i q [A_\mu , A_\nu]_*, \label{eomNC1}\\
A_{\nu}&:& \partial_\mu F^{\mu\nu} +i q [A_\mu ,F^{\mu\nu}]_* = 0 \label{eomNC2}.
\ea
So, we do recover our sourced equation (\ref{modification1}) as the e.o.m. for $A$, while the relation between $F$ and $A$ (\ref{eomA}), gets modified to (\ref{eomNC1}), which are the new e.o.m. for $F$. Because of the non-locality of the whole construction, it is instructive to explicitly verify that it’s consistent. As in the previous section, because the extra term \( j^\mu * A_\mu \) does not contain derivatives (on top of the ones in \(*\)-product, which could be removed) the current will be still the same as in (\ref{NLcurrent1}):
\[
j^\mu \equiv j_0^\mu = iq [A_\nu, F^{\nu\mu}]_*
\]
and it’s conserved on the e.o.m. (\ref{eomNC1},\ref{eomNC2}):
\ba\label{consistency}
\frac{1}{iq}\partial_\mu j^\mu
&=& \bigl[ \partial_\mu A_\nu, F^{\nu\mu} \bigr]_* + \bigl[ A_\nu, \partial_\mu F^{\nu\mu} \bigr]_* \nonumber \\
&\overset{(\ref{eomNC1},\ref{eomNC2})}{=}& \frac{1}{2} \bigl[ F_{\mu\nu} - iq [A_\mu, A_\nu]_*, F^{\nu\mu} \bigr]_* +  \bigl[ A_\nu, iq \bigl[ A_\mu, F^{\mu\nu} \bigr]_* \bigr]_* \nonumber\\
&=& \frac{iq}{2} \bigl[ \bigl[ A_\mu, A_\nu \bigr]_*, F^{\mu\nu} \bigr]_* + iq \bigl[ \bigl[ F^{\mu\nu}, A_\mu \bigr]_*, A_\nu \bigr]_*  \nonumber\\
&\overset{\text{Jacobi}}{=}& \frac{iq}{2} \biggl( -\bigl[ \bigl[ F^{\mu\nu}, A_\mu \bigr]_*, A_\nu \bigr]_* - \bigl[ \bigl[ A_\nu, F^{\mu\nu} \bigr]_*, A_\mu \bigr]_* \biggr) + iq \bigl[ \bigl[ F^{\mu\nu}, A_\mu \bigr]_*, A_\nu \bigr]_* \equiv 0 .\nonumber
\ea
So, (\ref{action1}) is the exact action leading to a consistent self-interacting noncommutative $U(1)$ theory:
\[
S_{\text{NC}} =
- \frac{1}{2} \int d^4x\, \Bigl[F^{\mu\nu} \bigl( \partial_\mu A_\nu - \partial_\nu A_\mu + iq [A_\mu, A_\nu]_* \bigr)
- \frac{1}{2} F^{\mu\nu} F_{\mu\nu}\Bigr]
\]
or solving for $F_{\mu\nu}$ from (\ref{eomNC1})
\ba\label{NCaction}
S_{\text{NC}} =
- \frac{1}{4} \int d^4x\, F^{\mu\nu} F_{\mu\nu} ,
\ea
where now \( F_{\mu\nu} \) is not independent and the only dynamical field is $A_\mu$.

As usual, by re-scaling \( A_\mu \mapsto \frac{1}{q} A_\mu \), we can re-absorb \( q \):
\[
S_{\text{NC}} =
- \frac{1}{4q^2} \int d^4x\, F^{\mu\nu} F_{\mu\nu},
\quad
F_{\mu\nu} = \partial_\mu A_\nu - \partial_\nu A_\mu + i[A_\mu, A_\nu]_* .
\]

Clearly, in this theory the global $U(1)$ symmetry has been ``enhanced'' to the noncommutative gauge symmetry\footnote{The verification is exactly the same as in the commutative non-abelian case, cf. the discussion at the end of the previous section. This is because for the $U(1)$ case, noncommutativity algebraically is just non-abelianity and, as was mentioned, the integral plays the role of the trace.}
\ba\label{gauge_nc_u1}
A_\mu \mapsto A^u_\mu = u * A_\mu * u^\dagger + i \partial_\mu u * u^\dagger .
\ea

The main point of this derivation is that we do not have to rely on the notion of the noncommutative $U(1)$, i.e., on noncommutative internal geometry, only space-time is noncommutative. As in the commutative case, the noncommutative $U(1)$ emerges as the consistency condition instead of being fundamental (note that the global $U(1)$ is not sensitive to noncommutativity of space-time). Where noncommutativity really plays a role is in the choice of the interaction. Because the free theory is the same in the commutative and noncommutative cases, it does not probe the noncommutativity of space-time. Only after the interaction is turned on, one can access its structure. Then the commutative choice of the current (which is zero in an abelian case) will leave our theory to be a free theory on the commutative space-time, while the choice of the non-local $U(1)$ current (\ref{NLcurrent1}), after running the consistency procedure, leads to the noncommutative $U(1)$ gauge theory, which lives on the Moyal space-time.

\subsection{Noncommutative non-abelian theories}\label{Non-abelian}

As we mentioned, noncommutative \( U(1) \) is already non-abelian, so, it seems, one might expect little difference between applying the Deser's method either to global \( U(1) \) or to some non-abelian group \( G \). However, two points, one physical and one mathematical, can be used to argue that the situation must be indeed different.

From the physical point of view, the two cases, abelian and non-abelian, are radically different: while in the commutative case abelian fields always remain free (again, we only consider theory without matter), non-abelian ones can be consistently made self-interacting. Turning on noncommutativity dramatically changes the situation for the abelian case, namely, it ceases to be free, which is a serious change in the dynamics. At the same time, the effect of noncommutativity for the non-abelian case is expected to lead just to some corrections to already non-trivial dynamics.

From the mathematical point of view, we should expect extra complications stemming from the extra ``non-abelianity'' of the star-product. Really, in the \( U(1) \) case the Lie algebra \( \mathfrak{u}(1) \) has just one generator, so \( [A,B]_{*} \) is automatically Lie algebra valued for any $\mathfrak{u}(1)$-valued elements of \( \mathcal{A }\). The situation radically changes in the non-abelian case: Now
\ba\label{problem}
[A_a T^a, B_b T^b]_{*} = (A_a * B_b) \, T^a T^b - (B_b * A_a) \, T^b T^a
= \frac{1}{2}[A_a, B_b]_* \{T^a , T^b\} + \frac{1}{2} \{ A_a , B_b\}_* [T^a, T^b],
\ea
where, as usual, $\{ \cdot , \cdot \}$ denotes an anticommutator and $\{ \cdot , \cdot \}_*$ - star-anticommutator.
Only the second term is Lie algebra valued and the first one makes sense only in the universal enveloping algebra.

Let us see in detail how this affects the construction. To be more specific, we will consider the cases of \(G = U(N) \) or \( G = SU(N) \) global symmetries. The starting point is exactly as in the commutative case, i.e., we start with a free theory of \( N^2 \) for \( U(N) \) and \( N^2 - 1 \) for \( SU(N) \) abelian fields, with the free dynamics given by the action (\ref{action_free}) and the equations of motion (\ref{eomA_comm},\ref{eomB_comm}). As we know, this theory is invariant under the global \( G \)-transformations (\ref{global}), and this leads to the local current (\ref{current}). To obtain a non-local current, we need to follow the same trick as in the noncommutative \( U(1) \) case, i.e., we have to formally promote \( \mathbf{u} \in (S)U(N) \) to be non-constant with entries from \( \mathcal{A} \),
\ba\label{g_local}
\mathbf{u} = e_{*}^{i \omega_a T^a},
\ea
where \( \omega_a \in \mathcal{A} \).

Using the Baker-Campbell-Hausdorff formula, one immediately sees that, due to (\ref{problem}), if $\mathbf{u}$ and $\mathbf{v}$ are of the form (\ref{g_local}), their $*$-product, $\mathbf{u}*\mathbf{v}$, is not, in general, of this form anymore. So, for non-constant $\omega$'s, the set of elements of the form (\ref{g_local}) does not, in general, form a group.

Nevertheless, if $\omega$ is constant, this is not a problem and because promoting $\omega$'s to be non-constants is just a trick to get the N\"{o}ther charge, we still can try, for the moment, proceed as in the commutative case, i.e., without specifying the representation of the generators $T^a$. The only requirement would be the same normalization as in (\ref{generators}). Then starting with the same free action (\ref{action_free}), invariant under the global symmetry (\ref{global})
\ba\label{action_free_NC}
S_0 = -\frac{1}{2}\int d^4 x\, \Tr\left[\mathbf{F}^{\mu\nu}(\partial_\mu \mathbf{A}_\nu - \partial_\nu \mathbf{A}_\mu) - \frac{1}{2} \mathbf{F}^{\mu\nu}\mathbf{F}_{\mu\nu}\right],
\ea
we end up with the same free dynamics (\ref{eomA_comm},\ref{eomB_comm}). Formally promoting $\mathbf{u}$ as in (\ref{g_local}), we immediately arrive at the non-local conserved current that looks like (\ref{NLcurrent1})
\ba\label{NLcurrent2}
\mathbf{j}_0^\mu = i q [\mathbf{A}_\nu , \mathbf{F}^{\nu\mu}]_*,\ \ \partial_\mu \mathbf{j}_0^\mu = 0\ \ \mathrm{on\ the\ e.o.m.}.
\ea
The conservation on the equations of motion does not depend on the choice of the representation. The crucial difference between (\ref{NLcurrent1}) and (\ref{NLcurrent2}) is that now it has the form (\ref{problem}):
\ba\label{NLcurrent3}
\mathbf{j}_0^\mu = \frac{i q }{2}[A_{a\nu}, F^{\mu\nu}_b]_* \{T^a , T^b\} + \frac{i q }{2} \{ A_{a\nu} , F^{\mu\nu}_b\}_* [T^a, T^b]
\ea
and, of course, this (namely, the first term in (\ref{NLcurrent3})) is a consequence of the fact discussed after Eq.(\ref{g_local}), that, in general, $\mathbf{u}$'s do not form a group under the $*$-multiplication.

If one wants, as before, to use (\ref{NLcurrent3}) as a source for the free equations of motion, one has to ``match'' the structures of $\mathbf{F}^{\nu\mu}$ and $\mathbf{j}_0^\mu$. But while $\mathbf{F}^{\nu\mu} = F^{\mu\nu}_a T^a$, $\mathbf{j}_0^\mu$ cannot be written as a linear combination of just the generators $T^a$ for a general algebra and a general representation (i.e., $ \mathbf{j}_0^\mu$ is not generally Lie-algebra-valued). For instance, in the case $\mathfrak{g} = \mathfrak{su}(N)$, if we want to take $T^a$ in the adjoint representation, as it is customary in the commutative case, we will have
\ba\label{tensor}
\mathbf{ad}\otimes \mathbf{ad} = \mathbf{1}\oplus \mathbf{ad} \oplus \mathbf{higher},
\ea
where higher means higher representations. The problem with (\ref{tensor}) is that the right hand side has more ``stuff'' in addition to the original generators in the adjoint representation, $\mathbf{1}$ and $\mathbf{higher}$, and this will contribute unwanted terms to the anticommutator in (\ref{NLcurrent3}).\footnote{For example, in the simplest $\mathfrak{su}(2)$ case, the adjoint representation is the same as the vector one, $\mathbf{ad} = \mathbf{3}$ and (\ref{tensor}) becomes $\mathbf{3}\otimes \mathbf{3} = \mathbf{1}\oplus \mathbf{3} \oplus \mathbf{5}$. The symmetric part of this is $\mathbf{1} \oplus \mathbf{5}$, which leads to the anticommutator $\{T^a , T^b\} = -\frac{4}{3}\mathds{1} + \mathbf{R}^{ab}$, where $(\mathbf{R}^{ab})_{cd} = -\frac{2}{3}\delta^{ab}\delta^{cd} + \delta^{ad}\delta^{bc} + \delta^{bd}\delta^{ac}$ is symmetric and traceless, i.e., IRR $\mathbf{5}$. This is a particular case of the general $\mathfrak{su}(N)$ in the adjoint representation: $\{T^a , T^b\} = -\frac{2N}{N^2 - 1}\mathds{1} + \mathbf{R}^{ab}$, where $\mathbf{R}$ contains symmetric traceless higher IRR's.} This can be solved in two ways: 1) we can admit from the start that $\mathbf{A}$ and $\mathbf{F}$ live in the universal enveloping algebra leading to an infinite number of free fields at the very first step of the procedure. We postpone the detailed study of this case and its relation to the known constructions of enveloping algebra valued fields \cite{Jurco:2000ja} to the future; 2) we should restrict ourselves to a very specific choice of the algebra in a very specific representation, namely when $\mathfrak{g} = \mathfrak{u}(N)$ and the generators are taken in the defining (which in this case is a fundamental) representation. This is, essentially, due to the Fierz-like identity, which is the reflection of the fact that $N\times N$ unitaries form a basis in $Mat_{N}$ and this leads to the following anticommutator in the defining representation (for the $\mathfrak{su}(N)$ part of the generators)
\ba\label{anticomm}
\{T^a , T^b\} = d^{abc}T^c + \frac{2}{N}\mathds{1},
\ea
where $d^{abc}$ is a totally symmetric tensor. This shows that only in the case of unitary groups in the defining representation the conserved current (\ref{NLcurrent3}) will be still Lie algebra valued (because $\mathds{1}$ is already part of their Lie algebra $\mathfrak{u}(N)$) and can be used as a source in the Deser's method. If one would start with $G=SU(N)$, from (\ref{anticomm}) we see that the conserved current would still generate the missing $U(1)$ part ``lifting'' $SU(N)$ to $U(N)$.

At this point, it should be clear that because $\int\,d^4 x$ plays a role of the trace on the algebra $(\mathcal{A}, *)$, after restricting to the $U(N)$ case and the defining representation, the rest of the consistency procedure will follow exactly the same steps as in the noncommutative $U(1)$ case. The full action for the consistent self-interacting noncommutative theory will be given by the $*$-analog of (\ref{action_inter})
\ba\label{action_inter_NC}
S = S_0 +\frac{1}{2}\int d^4 x\, \Tr \bigl( \mathbf{j}_0^\mu *\mathbf{A}_\mu \bigr) = -\frac{1}{2}\int d^4 x\, \Tr\left[\mathbf{F}^{\mu\nu}(\partial_\mu \mathbf{A}_\nu - \partial_\nu \mathbf{A}_\mu + iq [\mathbf{A}_\mu , \mathbf{A}_\nu]_* ) - \frac{1}{2} \mathbf{F}^{\mu\nu}\mathbf{F}_{\mu\nu}\right] ,
\ea
leading to the consistent self-interacting dynamics
\ba\label{eom_nc_nonab}
\mathbf{F}^{\mu\nu}&:& \mathbf{F}_{\mu\nu} = \partial_\mu \mathbf{A}_\nu - \partial_\nu \mathbf{A}_\mu + iq [\mathbf{A}_\mu , \mathbf{A}_\nu]_* , \label{eomA_comm2}\\
\mathbf{A}_{\nu}&:& \partial_\mu \mathbf{F}^{\mu\nu} = -i q [\mathbf{A}_\mu , \mathbf{F}^{\mu\nu}]_* \equiv -\mathbf{j}^\nu \label{eomB_comm2}.
\ea
In the same way as above, one can check the emergence of the noncommutative gauge symmetry
\ba\label{gauge_nc_u2}
\mathbf{A}_\mu \mapsto \mathbf{A}^u_\mu = \mathbf{u} * \mathbf{A}_\mu * \mathbf{u}^\dagger + \frac{i}{q} \partial_\mu \mathbf{u} * \mathbf{u}^\dagger .
\ea
It is very important to notice that the same condition on the group and the representation that was used above to construct a consistent self-interaction, guarantees that (\ref{gauge_nc_u2}) is well-defined, i.e., after a gauge transformation, $\mathbf{A}^u_\mu$ remains Lie algebra valued.

\section{Discussion and conclusion}

In this paper, we gave a non-geometric construction of noncommutative gauge fields on the Moyal space-time following the approach first introduced by Deser for the commutative case. We showed that insisting on remaining within Lie algebraic setting (i.e., without resorting to enveloping algebras), our construction leads to noncommutative gauge theories known from the standard approach, i.e., noncommutative $U(N)$ gauge theories in fundamental representation. In our approach, these restrictions on the allowed groups and representations come from matching the algebraic structures of the conserved non-local current and of the free fields used in the construction, rather than from the requirement that the star-product is a group product: the local non-commutative group arises only as an emergent consequence. The importance of this result is that it demonstrates the uniqueness of the theory in the same way as the non-geometric construction of a self-interacting masless spin-2 theory, starting with the free Fierz-Pauli theory in the flat space-time, unambiguously leads to General Relativity \cite{Feynman:1996kb,Deser:1969wk}.

From a more physical point of view, our construction explicitly demonstrates  the role of non-local interaction in probing the structure of the space-time: at the level of a free theory, one cannot tell between the commutative and noncommutative theory and only a choice of either local or non-local current to built a self-interacting theory will make this distinction. This is consistent with the observation made in \cite{Pinzul:2011di} about the energy-momentum tensor of a noncommutative scalar field theory. In that case, also a local as well as a non-local energy-momentum tensors exist. While the former one is a tensorial object with respect to the standard Poincare symmetry, the latter, non-local, is a tensor with respect to its twisted version, which can be thought as the actual symmetry of the Moyal space-time \cite{Chaichian:2004za}.

This brings the following question. In \cite{Balachandran:2006ib,Balachandran:2007kv,Balachandran:2007yf}, it was shown that (in quantum theory) it is possible to consistently couple a commutative gauge field to a noncommutative matter sector. It would be instructive to see what happens to such models in our approach.

More importantly, we consider the present work as an essential preparatory step towards the construction of noncommutative gravity. While there is more or less general consensus about noncommutative gauge fields, the construction of a natural noncommutative generalization of General Relativity is far from being settled, see \cite{Aschieri:2005zs,Harikumar:2006xf,Schenkel:2011biz,Aschieri:2022kzo} for some approaches. This is largely due to the fact that there is no unique approach to the Moyal-star-based noncommutative deformation of Riemannian geometry. Then, having a non-geometric construction, in which one starts with the same Fierz-Pauli theory as in the commutative case, and in which the noncommutative Riemannian geometry would be emergent, should be very useful.\footnote{Note, that in this case, the Moyal star would be much more natural, because one starts with the flat noncommutative background - the Moyal space-time. Introducing the Moyal star in a non-flat setting, as it is usually done in more naive approaches, is far from being natural.}

Another question to be addressed, briefly mentioned in the main text, is understanding of how the construction works in a general case, when one does not put the unitary restriction on groups and does not require the specific, fundamental, representation. From the discussion in subsection \ref{Non-abelian}, it is clear that one will have to extend the construction to enveloping algebra setting (even at the free level, one will have to allow an infinite number of fields!), but the details of this extension should be carefully worked out. In particular, a very important question on the uniqueness of this kind of models, i.e., whether our non-geometric construction will lead to the known noncommutative gauge models based on enveloping algebras, must be studied.

Some of the questions raised above are currently under study and the results will be reported elsewhere, the others and related we are planning to address in the nearest future.

\section*{Acknowledgements}

The authors have benefited from discussions with Carolina Gregory during various stages of the project and thank Allen Stern for clarifying comments on the final draft.  A.P. acknowledges the partial support of CNPq under the grant no.312842/2021-0.

\bibliographystyle{utphys}
\bibliography{Ref}

\providecommand{\href}[2]{#2}\begingroup\raggedright\begin{thebibliography}{10}

\bibitem{Snyder:1946qz}
H.~S. Snyder, ``{Quantized space-time},''
  \href{http://dx.doi.org/10.1103/PhysRev.71.38}{{\em Phys. Rev.} {\bfseries
  71} (1947) 38--41}.

\bibitem{Seiberg:1999vs}
N.~Seiberg and E.~Witten, ``{String theory and noncommutative geometry},''
  \href{http://dx.doi.org/10.1088/1126-6708/1999/09/032}{{\em JHEP} {\bfseries
  09} (1999) 032}, \href{http://arxiv.org/abs/hep-th/9908142}{{\ttfamily
  arXiv:hep-th/9908142}}.

\bibitem{Douglas:2001ba}
M.~R. Douglas and N.~A. Nekrasov, ``{Noncommutative field theory},''
  \href{http://dx.doi.org/10.1103/RevModPhys.73.977}{{\em Rev. Mod. Phys.}
  {\bfseries 73} (2001) 977--1029},
  \href{http://arxiv.org/abs/hep-th/0106048}{{\ttfamily arXiv:hep-th/0106048}}.

\bibitem{Szabo:2001kg}
R.~J. Szabo, ``{Quantum field theory on noncommutative spaces},''
  \href{http://dx.doi.org/10.1016/S0370-1573(03)00059-0}{{\em Phys. Rept.}
  {\bfseries 378} (2003) 207--299},
  \href{http://arxiv.org/abs/hep-th/0109162}{{\ttfamily arXiv:hep-th/0109162}}.

\bibitem{Jurco:2000ja}
B.~Jurco, S.~Schraml, P.~Schupp, and J.~Wess, ``{Enveloping algebra valued
  gauge transformations for nonAbelian gauge groups on noncommutative
  spaces},'' \href{http://dx.doi.org/10.1007/s100520000487}{{\em Eur. Phys. J.
  C} {\bfseries 17} (2000) 521--526},
  \href{http://arxiv.org/abs/hep-th/0006246}{{\ttfamily arXiv:hep-th/0006246}}.

\bibitem{Deser:1969wk}
S.~Deser, ``{Selfinteraction and gauge invariance},''
  \href{http://dx.doi.org/10.1007/BF00759198}{{\em Gen. Rel. Grav.} {\bfseries
  1} (1970) 9--18}, \href{http://arxiv.org/abs/gr-qc/0411023}{{\ttfamily
  arXiv:gr-qc/0411023}}.

\bibitem{Feynman:1996kb}
R.~P. Feynman, {\em {Feynman lectures on gravitation}}.
\newblock Addison-Wesley, 1995.

\bibitem{Hall:371445}
B.~C. Hall, {\em {L}ie {G}roups, {L}ie {A}lgebras, and {R}epresentations: {A}n
  {E}lementary {I}ntroduction}, vol.~222 of {\em Graduate Texts in
  Mathematics}.
\newblock Springer, New York, 2010.

\bibitem{Deser:1963zzc}
S.~Deser and R.~Arnowitt, ``{Interaction Among Gauge Vector Fields},''
  \href{http://dx.doi.org/10.1016/0029-5582(63)90081-6}{{\em Nucl. Phys.}
  {\bfseries 49} (1963) 133--143}.

\bibitem{Groenewold:1946kp}
H.~J. Groenewold, ``{On the Principles of elementary quantum mechanics},''
  \href{http://dx.doi.org/10.1016/S0031-8914(46)80059-4}{{\em Physica}
  {\bfseries 12} (1946) 405--460}.

\bibitem{Moyal:1949sk}
J.~E. Moyal, ``{Quantum mechanics as a statistical theory},''
  \href{http://dx.doi.org/10.1017/S0305004100000487}{{\em Proc. Cambridge Phil.
  Soc.} {\bfseries 45} (1949) 99--124}.

\bibitem{Pinzul:2011di}
A.~Pinzul, ``{UV/IR mixing as a twisted Poincar{\'e} anomaly},''
  \href{http://dx.doi.org/10.1088/1751-8113/45/7/075401}{{\em J. Phys. A}
  {\bfseries 45} (2012) 075401},
  \href{http://arxiv.org/abs/1109.5137}{{\ttfamily arXiv:1109.5137 [hep-th]}}.

\bibitem{Chaichian:2004za}
M.~Chaichian, P.~P. Kulish, K.~Nishijima, and A.~Tureanu, ``{On a
  Lorentz-invariant interpretation of noncommutative space-time and its
  implications on noncommutative QFT},''
  \href{http://dx.doi.org/10.1016/j.physletb.2004.10.045}{{\em Phys. Lett. B}
  {\bfseries 604} (2004) 98--102},
  \href{http://arxiv.org/abs/hep-th/0408069}{{\ttfamily arXiv:hep-th/0408069}}.

\bibitem{Balachandran:2006ib}
A.~P. Balachandran, A.~Pinzul, B.~A. Qureshi, and S.~Vaidya, ``{QED On the
  Groenewold-Moyal Plane},''
  \href{http://dx.doi.org/10.1142/S0217751X09046837}{{\em Int. J. Mod. Phys. A}
  {\bfseries 24} (2009) 4789--4804},
  \href{http://arxiv.org/abs/hep-th/0608138}{{\ttfamily arXiv:hep-th/0608138}}.

\bibitem{Balachandran:2007kv}
A.~P. Balachandran, A.~Pinzul, B.~A. Qureshi, and S.~Vaidya, ``{Twisted Gauge
  and Gravity Theories on the Groenewold-Moyal Plane},''
  \href{http://dx.doi.org/10.1103/PhysRevD.76.105025}{{\em Phys. Rev. D}
  {\bfseries 76} (2007) 105025},
  \href{http://arxiv.org/abs/0708.0069}{{\ttfamily arXiv:0708.0069 [hep-th]}}.

\bibitem{Balachandran:2007yf}
A.~P. Balachandran, A.~Pinzul, B.~A. Qureshi, and S.~Vaidya, ``{S-Matrix on the
  Moyal Plane: Locality versus Lorentz Invariance},''
  \href{http://dx.doi.org/10.1103/PhysRevD.77.025020}{{\em Phys. Rev. D}
  {\bfseries 77} (2008) 025020},
  \href{http://arxiv.org/abs/0708.1379}{{\ttfamily arXiv:0708.1379 [hep-th]}}.

\bibitem{Aschieri:2005zs}
P.~Aschieri, M.~Dimitrijevic, F.~Meyer, and J.~Wess, ``{Noncommutative geometry
  and gravity},'' \href{http://dx.doi.org/10.1088/0264-9381/23/6/005}{{\em
  Class. Quant. Grav.} {\bfseries 23} (2006) 1883--1912},
  \href{http://arxiv.org/abs/hep-th/0510059}{{\ttfamily arXiv:hep-th/0510059}}.

\bibitem{Harikumar:2006xf}
E.~Harikumar and V.~O. Rivelles, ``{Noncommutative Gravity},''
  \href{http://dx.doi.org/10.1088/0264-9381/23/24/024}{{\em Class. Quant.
  Grav.} {\bfseries 23} (2006) 7551--7560},
  \href{http://arxiv.org/abs/hep-th/0607115}{{\ttfamily arXiv:hep-th/0607115}}.

\bibitem{Schenkel:2011biz}
A.~Schenkel, {\em {Noncommutative Gravity and Quantum Field Theory on
  Noncommutative Curved Spacetimes}}.
\newblock PhD thesis, Wurzburg U., 2011.
\newblock \href{http://arxiv.org/abs/1210.1115}{{\ttfamily arXiv:1210.1115
  [math-ph]}}.

\bibitem{Aschieri:2022kzo}
P.~Aschieri and L.~Castellani, ``{Noncommutative gauge and gravity theories and
  geometric Seiberg{\textendash}Witten map},''
  \href{http://dx.doi.org/10.1140/epjs/s11734-023-00831-7}{{\em Eur. Phys. J.
  ST} {\bfseries 232} no.~23-24, (2023) 3733--3746},
  \href{http://arxiv.org/abs/2209.03774}{{\ttfamily arXiv:2209.03774
  [hep-th]}}.

\end{thebibliography}\endgroup

\end{document}